\documentclass[a4paper,10pt,oneside,onecolumn,headings=small]{scrartcl}

\usepackage[T1]{fontenc}
\usepackage{lmodern}
\usepackage{microtype}
\usepackage[margin=3cm]{geometry}
\usepackage{graphicx}
\usepackage{amsmath,amsfonts}
\usepackage{booktabs}
\usepackage{caption}
\usepackage{subcaption}
\usepackage{siunitx}
\usepackage{natbib}
\usepackage{xcolor} 
\definecolor{darkgreen}{rgb}{0.0,0.4,0.0}
\usepackage[colorlinks=true,
            urlcolor=blue,
            citecolor=blue,
            linkcolor=black]{hyperref}
\usepackage{siunitx}
\usepackage[affil-it]{authblk} 
\KOMAoptions{DIV=14}

\setkomafont{title}{\sffamily\bfseries\large}
\setkomafont{author}{\sffamily\normalsize}

\title{Operational machine learning for park-scale irrigation to support urban cooling}

\author{Mesut Koçyiğit\thanks{Current affiliation: NSW Department of Planning and Environment, Parramatta NSW 2150, Australia}}
\author{Bahman Javadi}
\author{Russell Thomson\thanks{Current affiliation: The Analytical Edge, Blackmans Bay, Tasmania 7052, Australia}}
\author{Sebastian Pfautsch}
\author{Oliver Obst\thanks{Corresponding author: \texttt{o.obst@westernsydney.edu.au}}}

\affil{Western Sydney University, NSW 2751, Penrith, Australia}

\date{} 

\begin{document}
\maketitle

\vspace*{-1cm}
\begin{abstract}
Urban parks can mitigate local heat, yet irrigation control is usually tuned for water savings rather than cooling. We report on SIMPaCT (Smart Irrigation Management for Parks and Cool Towns), a park-scale deployment that links per-zone soil-moisture forecasts to overnight irrigation set-points in support of urban cooling. SIMPaCT ingests data from 202 soil-moisture sensors, 50 temperature-relative humidity (TRH) nodes, and 13 weather stations, and trains a per-sensor k-nearest neighbours (kNN) predictor on short rolling windows (200-900\,h). A rule-first anomaly pipeline screens missing and stuck-at signals, with model-based checks (Isolation Forest and ARIMA). When a device fails, a mutual-information neighbourhood selects the most informative neighbour and a small multilayer perceptron supplies a ``virtual sensor'' until restoration. Across sensors the mean absolute error was 0.78\%, comparable to more complex baselines; the upper-quartile error (P75) was lower for kNN than SARIMA (0.71\% vs 0.93\%). SIMPaCT runs daily and writes proposed set-points to the existing controller for operator review. 
This short communication reports an operational recipe for robust, cooling-oriented irrigation at city-park scale.
\end{abstract}


\section{Motivation and contribution}

Urban green spaces are an important tool for mitigating the urban heat island effect~\citep{Pfa22,ATA24} and require careful environmental monitoring to maintain their ecological and social benefits~\citep{Pfa22,GWZ+25}. Large parks act as natural air conditioners; however, irrigation systems are often designed for water conservation rather than cooling.
We address the operational gap with SIMPaCT (Smart Irrigation Management for Parks and Cool Towns), by linking short-horizon soil-moisture forecasts to overnight irrigation set-points at zone level in a live 40 ha park (Bicentennial Park, Sydney Olympic Park). Recent microclimate measurements in the same precinct found a mean park--urban air-temperature differential of about $-0.4\,^{\circ}\mathrm{C}$ across summer 2022/23, with frequent night-time cooling exceeding $5\,^{\circ}\mathrm{C}$ and daytime cooling of approx. 1-2$^{\circ}\mathrm{C}$ on hot days \citep{Pfautsch2023PCI}.

\paragraph{Related work in brief}
Most ML-irrigation studies optimise water use rather than cooling \citep{nosratabadi2019state,ML_REVIEW}. Experimental field systems often couple small multi-sensor sets with classical learners such as kNN, logistic regression, SVM, naïve Bayes, and simple neural nets \citep{TTE+22,Rajan2023AIP,9156026,10.1155/2021/6691571}. Others use SVR or clustering over humidity, temperature and soil-moisture inputs \citep{10.3844/jcssp.2020.355.363,VIJ20201250}. Deep sequence models have been trialled in orchards or across a few sites \citep{agriculture12010025,9388691}, and ensemble regressors on mixed soil-air variables with UV radiation have also been reported \citep{8988313}. These systems are typically small-scale, agriculture-focused, or lab/experimental. We report a live, park-scale pipeline that retrains daily from short windows, integrates with legacy controllers, and remains operable under device faults via an information-theoretic “virtual sensor”.

\noindent\textbf{Contributions.} (i) A per-sensor kNN predictor trained from one week of data provides useful next-day soil-moisture forecasts. (ii) A hybrid anomaly pipeline combines simple rules with model-based checks. (iii) An information-theoretic neighbourhood selects a backup source so a small multilayer perceptron (MLP) can act as a temporary “virtual sensor” when devices fail. Forecasts of  per-zone proposals are visualised and can be accepted or overridden by operators. 
This short communication focuses on the operational ML layer: forecasting, anomaly screening, and the information-theoretic backup. System architecture, dashboards, and the broader digital-twin workflow are documented in the companion paper \citep{CUP+23}.

\section{System in brief}
\begin{figure}[tb]
  \centering
  \includegraphics[width=.495\linewidth]{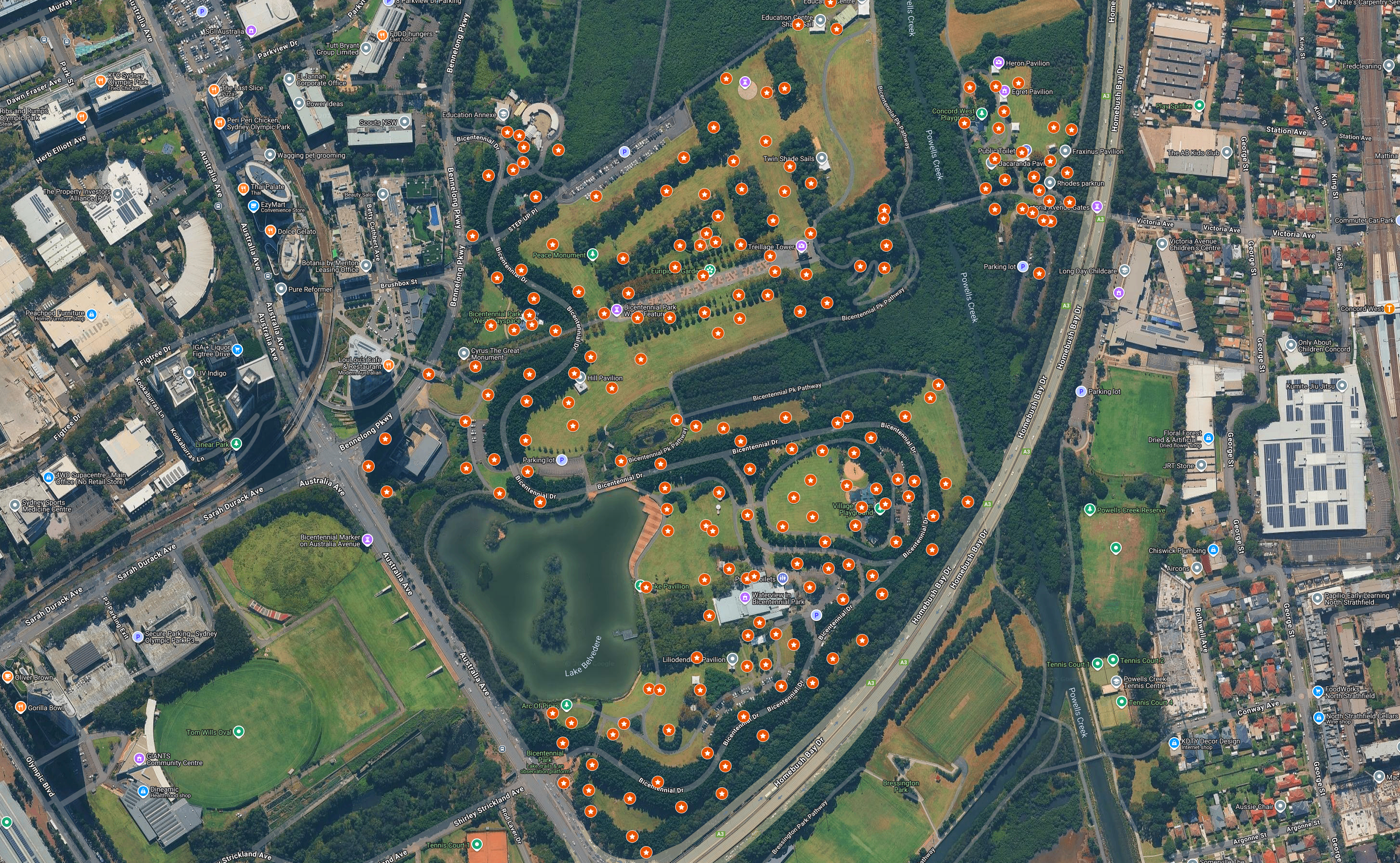}\hfill
  \includegraphics[width=.495\linewidth]{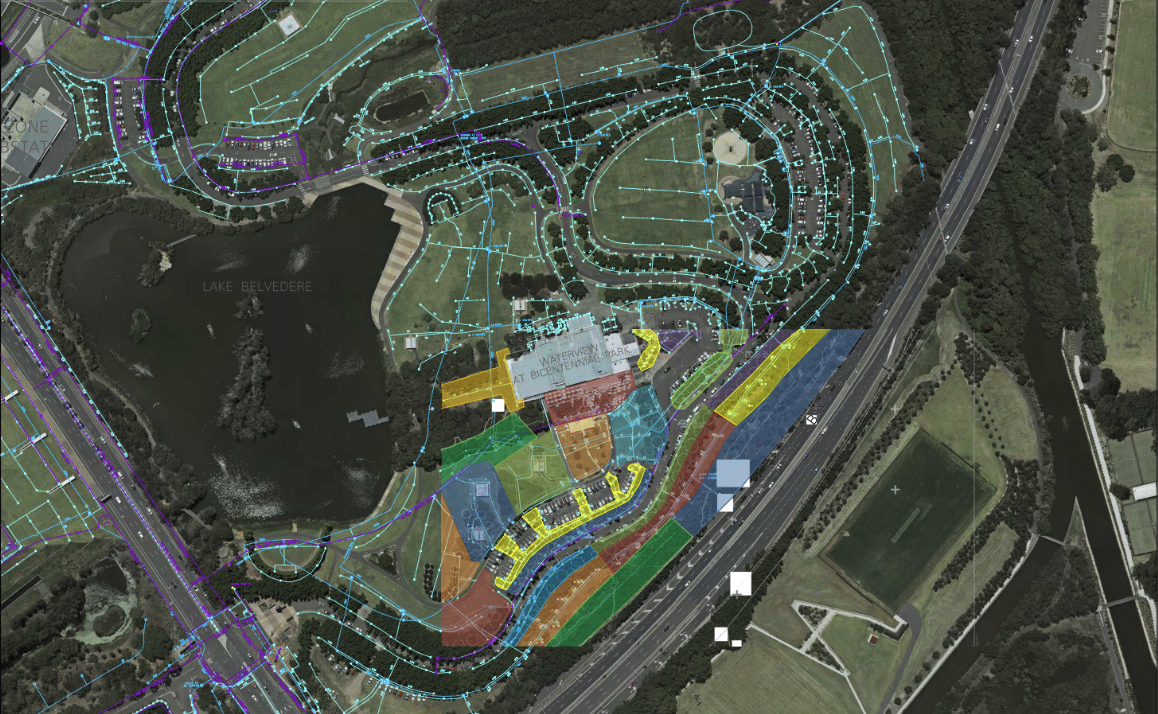}
  \caption{Park context: sensor placement (left, orange); part of the irrigation network (right, blue lines).}
  \label{fig:context}
\end{figure}

SIMPaCT operates 202 soil-moisture sensors, 50 temperature-relative humidity (TRH) sensors and 13 weather stations that transmit data via LoRaWAN (Long Range Wide Area Network)/4G. Weather stations were mounted on lamp posts at $\sim$3 m with short outriggers, sampled at 15~min cadence, and forwarded via 4G to the Fieldmouse platform; the deployment protocol and siting are detailed in \citep{Pfautsch2023PCI}. The SIMPaCT platform, built on Senaps~\citep{mac2017senaps}, ingests sensor streams and Bureau of Meteorology forecasts, maintains a per-zone digital twin and produces forecast soil moisture and proposed overnight runtimes. Proposals are posted to the existing irrigation network controller that operates the about 200 controllable sprinkler stations; accepted runtimes are logged. Figure \ref{fig:context} shows the station layout and network. A companion paper documents the digital-twin architecture, dashboards and end-to-end workflow in more detail \citep{CUP+23}.

\paragraph{Operational constraints and integration}
Irrigation is constrained by public access, pressure management and event blackouts. Only one main line runs at a time, watering in short overnight windows. SIMPaCT writes per-zone set-points to the legacy irrigation network controller and logs executed runtimes. Catch-can field tests provided runtime-to-application-rate factors so scheduled minutes can be compared with forecast precipitation on a common mm\,h$^{-1}$ scale.

\subsection*{Virtual sensor and backup predictor}
\begin{figure}[tb]
  \centering
  \includegraphics[width=0.95\linewidth]{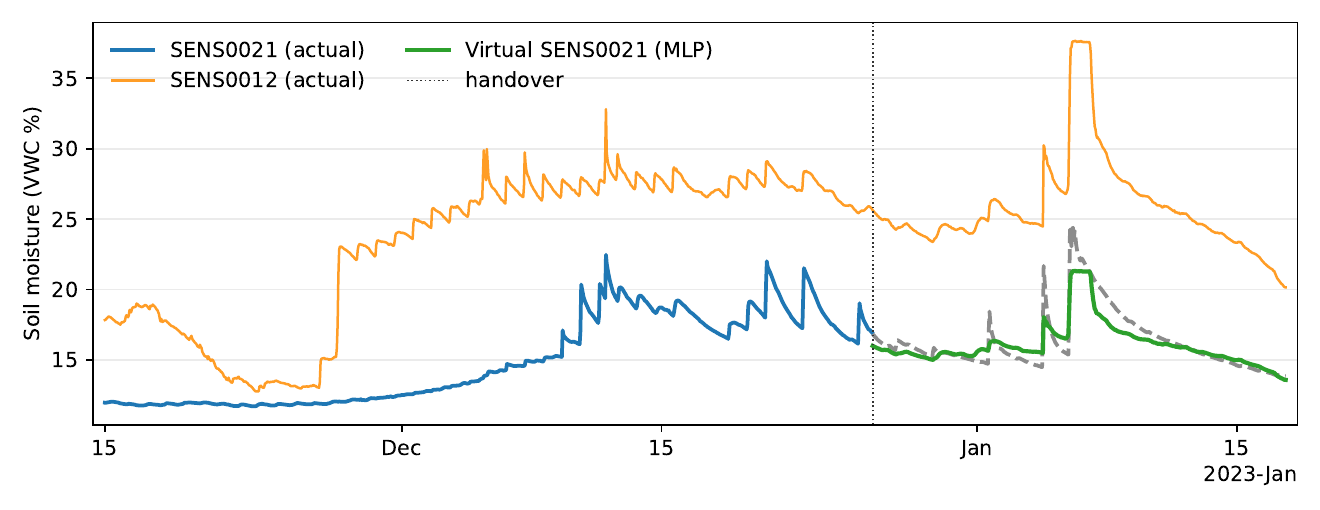}
  \caption{Backup during a simulated sensor failure (soil moisture, \%). The virtual sensor (MLP, 1×10 ReLU) is trained on early data to map SENS0012 (neighbour, the top line in orange) to SENS0021 (target, in blue), and supplies values (green, true values in grey, dashed) during the outage window, right of the vertical line. Streams are volumetric water content (VWC) resampled at hourly cadence.}
  \label{fig:backup}
\end{figure}

Sensor faults are inevitable in long-running deployments~\citep{JWOV11}.
When a sensor is flagged faulty by the rules or model checks, SIMPaCT activates a temporary ``virtual sensor''. We precompute an information-theoretic neighbourhood by estimating pairwise mutual information (MI) between the continuous sensor time series using the Kraskov--Stögbauer--Grassberger $k$-NN estimator as implemented in JIDT~\citep{KSG04,Liz14}. We use the standard definition of mutual information (natural logs):
\begin{equation}
I(X;Y) \;=\; \mathbb{E}_{p_{XY}} \!\left[\log \frac{p_{XY}(x,y)}{p_X(x)\,p_Y(y)}\right].
\label{eq:mi}
\end{equation}
This yields a connectivity matrix; for each target we retain the single most-informative neighbour (top-1 by $I$, Eq.~\ref{eq:mi}). On activation, a small multilayer perceptron (one hidden layer, 10 ReLU units) is trained on the recent history (200--900\,h) to map the neighbour's series to the target. The virtual sensor outputs values at the same cadence as the failed device and is refreshed daily until the physical sensor returns. This keeps the architecture simple, avoids chaining many dependencies, and limits drift, while providing continuity for forecasting and set-point generation. The MI matrix is computed offline and refreshed infrequently; at runtime this paper’s contribution is the MI-selected neighbour plus a small MLP backup predictor that maps the neighbour’s series to the target, while visualisation and operator workflow are handled by the platform described in \citep{CUP+23}. For Fig.~\ref{fig:backup} and Fig.~\ref{fig:nimap} we used a 13-sensor subset resampled to hourly cadence from device readings between 2022-11-15 and 2023-01-17~\citep{KJT+25}; in production the MI matrix covers the full network.

\paragraph{Neighbour choice}
We retain only the most informative neighbour per target (top-1 by MI). This keeps the backup graph sparse, limits cascading dependencies if another device fails, and allows a small MLP to train quickly on the fly. In practice, a monthly MI matrix recomputation (or after topology changes) was sufficient because inter-sensor relationships were stable.

\subsection*{Rule-based anomaly checks}
We apply simple rules before model-based detectors:
\begin{itemize}
\item[-] \textbf{Missingness}: more than 50\,\% missing values within the training window.
\item[-] \textbf{Long gaps}: consecutive missing values over 3 days.
\item[-] \textbf{Stuck-at}: absolute change below 1\,\% over 120 hours.
\item[-] \textbf{Out-of-range}: values outside the sensor’s valid range.
\item[-] \textbf{Spikes}: extreme deviations relative to a 24\,h moving median.
\end{itemize}
These rules avoid use of corrupted data segments and trigger the virtual backup predictor when needed.

\subsection*{Model-based anomaly checks}
After rule screening we run two lightweight detectors on stationary series. 
(i) \textit{Isolation Forest} on weekly/seasonally differenced values (200 trees, contamination $=0.05$) flags a sensor if more than 30\,\% of samples in the window are classified anomalous~\citep{LTZ08}. 
(ii) A 24\,h \textit{ARIMA} forecast on seasonally differenced data raises a flag when $|$actual$-$forecast$| > 2\sigma$ for more than 30\,\% of points. 
Either flag activates the virtual sensor until the device is restored.

\section{Findings and operations}
\begin{figure}[tb]
  \centering
  \includegraphics[width=0.9\linewidth]{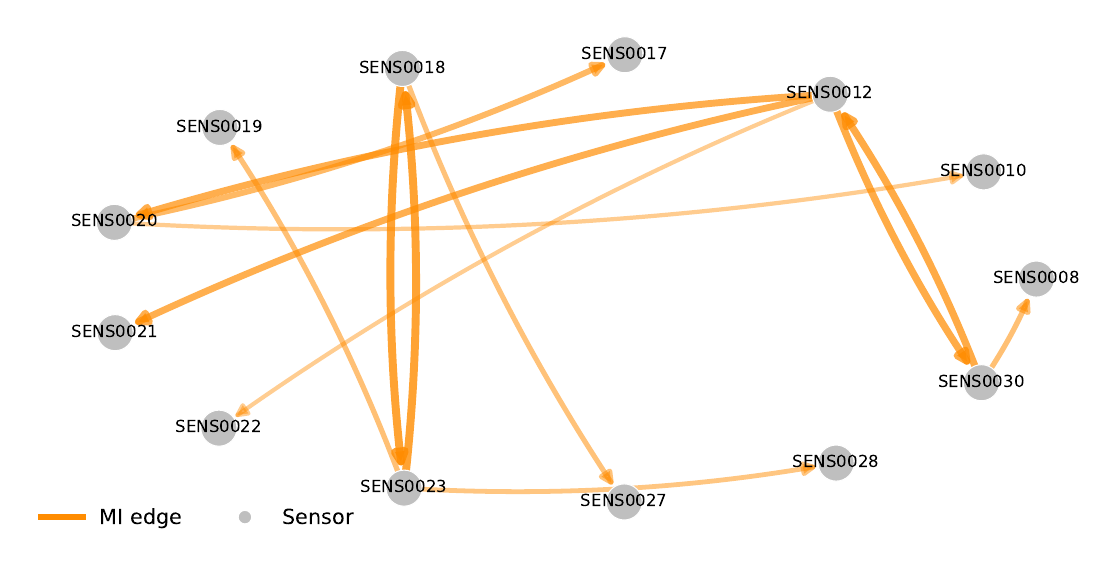}
    \caption{Mutual-information neighbourhood used to select backup source, derived from a 13-sensor test subset (hourly VWC, 2022-11-15 to 2023-01-17). Line width scales with MI.}
  \label{fig:nimap}
\end{figure}

Across sensors the mean absolute error (MAE) was 0.78\,\%. SARIMA achieved 0.73\,\% mean MAE, but kNN had lower upper-quartile error (P75 0.71\,\% vs 0.93\,\%) and retrained in minutes with minimal tuning. Daily operation runs at 16:00 local time: ingest latest data and forecasts, refresh models, generate 24-72 h forecasts, apply anomaly checks and activate backups, then post per-zone runtimes for overnight execution. Operators can accept or override proposals. In addition to the modelling results reported here, operational outcomes from the public SIMPaCT final report \citep{PW24} document day-to-day scheduling at precinct scale, controller integration logs, and maintenance workflows, including examples where the virtual sensor maintained continuity during device outages and case studies of rapid schedule adjustments ahead of forecast hot spells \citep{PW24}. These deployment notes complement this short communication by detailing dashboards, station-group water balances and operator review patterns over multiple seasons. Figure \ref{fig:backup} illustrates the fallback during a simulated failure, while Figure \ref{fig:nimap} shows the mutual-information neighbourhood used to select the backup source. The companion paper also describes hydrological modelling, a multi-objective optimiser targeting water efficiency or cooling, and an evapotranspiration fallback mode for sparse-data zones \citep{CUP+23}.

\section{Implications and limits}
Because park--urban differentials intensify in late afternoon and at night \citep{Pfautsch2023PCI}, the daily 16:00 scheduling window targets the periods when additional evapotranspiration can most plausibly amplify cooling. The PCI study also links higher hard-surface fractions to warmer air; our control complements structural interventions by timing irrigation to provide evaporative relief during hot spells.
The workflow functions as an operations-level planning support tool for heat mitigation, enabling same-day adjustments during hot spells and integration with park events and pressure constraints. Limitations include a focus on short-horizon prediction rather than quantified cooling or water savings, dependence on forecast quality and logs of executed runtimes, and reduced backup quality in sparse networks. Transfer to other parks may require re-tuning the window length, anomaly thresholds and runtime-to-application-rate conversion.

\newpage
\section*{CRediT authorship contribution statement}
\noindent
\textbf{Mesut Koçyiğit}: Investigation, Software, Writing-original draft. \\
\textbf{Bahman Javadi}: Conceptualisation, Validation, Formal analysis, Investigation, Writing-review \& editing, Funding acquisition. \\
\textbf{Russell Thomson}: Methodology, Validation, Formal analysis, Investigation, Writing-review \& editing. \\
\textbf{Sebastian Pfautsch}: Conceptualisation, Validation, Investigation, Resources, Data Curation, Writing-review \& editing, Supervision, Project administration, Funding acquisition. \\
\textbf{Oliver Obst}: Conceptualisation, Methodology, Validation, Formal analysis, Investigation, 
Writing-Original Draft, Writing-review \& editing, Supervision, Funding acquisition.

\section*{Declaration of competing interest}
The authors declare no competing interests.

\section*{Funding}
This work was partially funded by the NSW Government’s Digital Restart Fund (Smart Places/Smart Cities Acceleration Program), with co-funding and in-kind support from Sydney Olympic Park Authority and Sydney Water.

\section*{Data availability} 
The dataset supporting Figures 2 and 3 contains hourly volumetric water content (VWC, \%) for 13 soil-moisture sensors in Bicentennial Park, recorded between 2022-11-15 00:00 and 2023-01-17 15:00. It was derived from 118,024 raw sensor readings, aggregated to 1,528 hourly time slots with no missing values. Raw and processed data, along with preprocessing scripts and outputs for generating the mutual-information neighbourhood graph, are publicly available under a CC BY-4.0 license for data and an MIT license for code at \url{https://doi.org/10.5281/zenodo.16811020}.

\bibliographystyle{elsarticle-num-names}
\bibliography{definitions,simpact-irrigation}

\end{document}